\documentclass[aps, twocolumn, prb, showpacs,preprintnumbers,amsmath,amssymb,superscriptaddress]{revtex4-1}
\usepackage{amsmath,amsfonts,amssymb,graphics,graphicx,epsfig,color,times,indentfirst,layout,hyperref,subfigure,multirow,bm}
\usepackage{hyperref}
\usepackage{bbold}
\usepackage{ulem}
\hypersetup{linktocpage,,colorlinks=true,citecolor=red}
\usepackage{soul}

\newcommand{\be}{\begin{equation}}
\newcommand{\een}{\end{equation*}}
\newcommand{\bs}{\begin{split}}
\newcommand{\ben}{\begin{equation*}}
\newcommand{\ee}{\end{equation}}
\newcommand{\es}{\end{split}}
\newcommand{\bmx}{\begin{array}}
\newcommand{\emx}{\end{array}}
\newcommand{\bea}{\begin{eqnarray}}
\newcommand{\bean}{\begin{eqnarray*}}
\newcommand{\eea}{\end{eqnarray}}
\newcommand{\eean}{\end{eqnarray*}}
\newcommand{\dg}{^{\dagger}}
\newcommand{\dn}{^{\vphantom{\dagger}}}

\newcommand{\ua}{\uparrow}
\newcommand{\da}{\downarrow}
\newcommand{\Ua}{\Uparrow}
\newcommand{\Da}{\Downarrow}
\newcommand{\bb}[1]{\mathbb{#1}}
\newcommand{\qqquad}{\qquad\qquad\qquad}

\newcommand{\eps}{\epsilon}

\newcommand{\sgn}[1]{{\rm sign}{#1}}
\newcommand{\pref}[1]{(\ref{#1})}

\newcommand{\im}[1]{{\rm Im}\left[ #1 \right]}

\newcommand{\abs}[1]{\left\vert #1 \right\vert}

\newcommand{\ket}[1]{\left\vert #1\right\rangle}
\newcommand{\braket}[1]{\left\langle #1\right\rangle}

\newcommand{\mat}[1]{\left(\bmx{cc}#1\emx\right)}

\newcommand{\matl}[1]{\bmx{ll}#1\emx}

\newcommand{\bw}[1]{\begin{widetext}}
\newcommand{\ew}[1]{\end{widetext}}

\newcommand{\gray}[1]{}%{{\small \color{gray} #1}}

\begin{document}
\title{Renormalized dispersing multiplets in the spectrum of nearly Mott localized systems}
\author{Yashar Komijani}
\email{komijani@physics.rutgers.edu}
\affiliation{Department of Physics and Astronomy, Rutgers University, Piscataway, New Jersey, 08854, USA}

\author{Karen Hallberg}
 \affiliation{Centro At\'omico Bariloche and Instituto Balserio, CNEA and CONICET 8400 Bariloche, Argentina}

\author{Gabriel Kotliar}
\affiliation{Department of Physics and Astronomy, Rutgers University, Piscataway, New Jersey, 08854, USA}

\date{\today}
\begin{abstract}
{The spectrum of the strongly correlated systems usually shows resonant peaks at finite energy, with examples in the 115 Ce family which are reproduced by the dynamical mean-field theory. A similar structure has been seen recently in the orbitally selective Mott (OSM) phase of two-band model, known as doublon-holon bound state, with implications on the fate of such phase in the zero Hund's coupling limit. We show that these features can be captured with the slave-particle methods once their Hilbert space is taken into account. We use slave-spin calculations, justifiable in the limit of large dimensions, to explicitly demonstrate this and compare the results with dynamical mean-field theory.}\end{abstract}
\maketitle
\section{Introduction}
The study of the properties of quantum materials, including high-temperature superconductors, requires understanding 
strongly correlated systems in two or three dimensions, a task which is theoretically very challenging. Dynamical mean-field theory (DMFT),\cite{Georges96} one of the very few tools at our disposal, provides a systematic interpolation between the atomic limit where the correlations are important and the non-interacting limit of band theory, and is exact in the limit of infinite dimensions. However, analytical insight into the result is often formidable as a result of the self-consistency loop. 

There have been lots of efforts to produce analytically tractable understanding using slave-particle mean-field theories.\,\cite{Kotliar86,Lechermann07,
Medici05,Koga05,Winograd14,Yu13} While when treated exactly all these methods in principle agree, the approximation schemes used for analytical/numerical tractability causes discrepancies, for example on the fate of the so-called orbitally selective Mott (OSM) phase {in absence of Hund's coupling}, that has been a source of confusion. The simplest version of the phenomena appears in a two-band Hubbard model with orbital-dependent tunnelling and local interaction. Slave spin methods\,\cite{Medici05,Komijani17} predict that when the ratio of the bandwidths of the two bands $r=t_2/t_1$ is close to one, the two bands undergo a transition between Mott-phase and the metalic phase at the same value of Hubbard $U$, the so-called {\it locking effect}, whereas when the bandwidth ratio is smaller than a threshold of about $r_c=0.2$, there is a region of OSM phase in the phase diagram, in which one band is metalic and the other band is itinerant. This agrees with some\,\cite{Medici05} and disagrees with other DMFT calculations.\,\cite{Koga05,Winograd14} 
 Moreover, the general consensus is that Hund's coupling $J_H$ favours OSM phase and decreases $r_c$. 

In a recent study, {some of us}\cite{Nunez18,Nunez18b} used a density-matrix renormalization group (DMRG) impurity solver \cite{Garcia04} to obtain very accurate DMFT results on the two-orbital problem. The existence of an OSM phase was obtained by previous work.
\,\cite{Medici05} In the absence of a Hund's interaction, N\'u\~nez Fern\'andez et al\,\cite{Nunez18} found locking, irrespective of the ratio of the bandwidths. Moreover, when including an inter-orbital Coulomb interaction $U_{12}$, they identified a resonant feature in the spectral function of the localized orbital in the OSM phase, dubbed  {\it holon-doublon} excitonic peak corresponding to a virtual bound state at energy scales of about $\Delta=U-U_{12}$. 

There have been previous studies of holon-doublon peaks in the literature\,\cite{Zhou14,Lee17,Lee17b} of the single-band Hubbard model where, due to the limited size of the local Hilbert space, the holon-doublon boundstate necessarily forms between nearby sites. In contrast, the two-site, two-orbital model provided in\,\cite{Nunez18} {with on-site interaction} shows that their holon-doublon boundstate forms in the two orbitals of the same site {and in this case they form well-defined quasiparticle peaks.}

The problem of understanding the origin of finite-energy multiplets in the spectrum of the strongly correlated quantum materials is general and not limited to the two-orbital case mentioned above. Here, we show that slave-particle mean-field methods are fully capable of capturing these finite-energy spectral features, and in particular, the holon-doublon peak. 
 Similar methods has been applied in the past to analyze the spectrum of mixed valence compounds, including Pu pnictides and chalcogenides\,\cite{Yee10} and the 115 Ce family: CeIrIn$_5$, CeCoIn$_5$ and CeRhIn$_5$.\,\cite{Haule10}

The structure of the paper is the following: In section II, we describe the general formalism of the method, as well as various approximation schemes. %Section III contains a numerical renormalization group comparison between the original and the slave-spin representations of the Anderson impurity problem. 
Section III applies the general method of section II to study the spectral functions of single and two-band lattices, including the holon-doublon bound state, using $Z_2$ slave-spin method. In section IV we compare numerical results from slave-spin to DMFT. Appendix A contains a comparison between exact diagonalization and slave-spin method applied to the two-orbital two-site problem. Appendix B and C contain diagonalization of the slave-spin Hamiltonian, and the spectral representation of slave-particle Green's function, respectively.

\section{General formalism}
In slave-particle methods we introduce a parton construction for the fermionic operator. $d_\alpha=\hat z_\alpha f_\alpha$, where $\hat z$ has its own Hilbert space (without loss of generality we restrict our discussion to the simpler cases in which $z$ and $f$ share the same index). As a result the Hilbert space is enlarged to the tensor product of that of the $f$ and $z$ particles. A constraint, usually imposed on averaged via few Lagrange multipliers, ensures that the averaged physical parameters are computed in the originally restricted part of the extended Hilbert space. As a next step
\be
H=\sum_{\braket{ij}\alpha\beta}t_{\alpha\beta}d\dg_{i\alpha}d\dn_{j\beta}+{\sum_i H_{int}}[i]
%\{d_{i\alpha},d\dg_{i\alpha}\}}
\ee
is mean-field decoupled into $H_{MF}=H_f+H_z$
where
\be
H_z=\sum_{\braket{ij}\alpha\beta}J^{\alpha\beta}_{ij}{\hat z\dg_{i\alpha}\hat z\dn_{j\beta}}+\sum_i H_{int}[i]\label{eq2}
%\{\hat z_{i\alpha}\}
\ee
and
\be
H_f=\sum_{\braket{ij}\alpha\beta}\tilde t_{ij}^{\alpha\beta}f\dg_{i\alpha}f\dn_{j\beta}
\ee
with the parameters given by
\be
\tilde t_{ij}^{\alpha\beta}=t_{ij}^{\alpha\beta}\braket{\hat z\dg_{i\alpha}z\dn_{j\beta}},\quad J_{ij}^{\alpha\beta}=t_{ij}^{\alpha\beta}\braket{f\dg_{i\alpha}f\dn_{j\beta}}
\ee
Here, $H_f$ describes the renormalized fermionic bands, whereas $H_z$ describes renrormalized atomic structure. The real advantage is that the constraints can be used to absorb $H_{int}[i]$ entirely in $H_z$. The ordered (disordered) phases of the $z$-lattice, are usually associated with itinerant (Mott-localized) phases of the original electrons.

The mean-field decoupling mentioned above neglects fluctuations in time that couple $H_f$ and $H_z$, whereas these fluctuations are captured in DMFT. On the other hand long-wavelength spatial fluctuations are present in Eq.\,\pref{eq2} and absent in single-site DMFT. 
In practice, however, in order to solve the resulting interacting Hamiltonian in $H_z$ one uses a single-site approximation, so that $H_z$ is transformed to
\be
H_{z,ss}=\sum_\alpha (h_\alpha \hat z\dg_\alpha+h.c.)+H_{int}\label{eq4}
\ee
The spatial correlations are lost in this process, similar to the single-site DMFT. Systematic approaches to improve this result has been achieved by i) cluster extensions,\,\cite{Hassan10,Lechermann07} or ii) long-wavelength magnon-type excitations.\,\cite{Ruegg10}

An energetic competition between these $H_f$ and $H_{z,ss}$ leads to itineracy or localization. Calculating the Green's function $G_d(\tau)=\braket{-Td_\alpha (\tau)d_\alpha\dg}$, using the assumption of the mean-field decoupling between $z$-s and $f$-s, we can write
\be
G_{d,\alpha\beta}(n,\tau)=\Pi_{\alpha\beta}(n,\tau)G_{f,\alpha\beta}(n,\tau)\label{eq6}
\ee
where $G_{f,\alpha\beta}(n,\tau)=\langle-Tf\dn_{n\alpha}(\tau)f\dg_{0\beta}\rangle$ and $\Pi_{\alpha\beta}(n,\tau)=\langle T\hat z_{n\alpha}(\tau)\hat z\dg_{0\beta}\rangle$. Defining $z_\alpha=\braket{\hat z_\alpha}$ and $Z_{\alpha\beta}=z_\alpha z^*_\beta$,  within single-site approximation 
\be
\Pi_{\alpha\beta}(n,\tau)=
\left\{
\matl{
Z_{\alpha\beta} & n\neq 0 \\ \Pi_{\alpha\beta}(\tau) & n=0
}
\right.\label{eq7}
\ee
Without lack of generality, in the following we restrict the discussion to the diagonal $\alpha=\beta$ elements. The function $\Pi_{\alpha\alpha}(\tau)$ ``knows'' about the renormalized atomic physics as seen from its spectral representation
\bea
\Pi_{\alpha\alpha}(i\nu_n)=e^{-\Omega/T}\sum_{nm}\frac{e^{-E_n/T}-e^{-E_m/T}}{i\nu_n+E_n-E_m}\abs{\braket{n\vert \hat z_\alpha\vert m}}^2\quad\label{eq8}
\eea
where $T$ is the temperature, $\nu_n=2\pi nT$ are Matsubara frequencies and $\Omega$ is the Gibbs free energy, here. This atomic structure in $\Pi(\tau)$ is reflected in $G_d$ after convolution with the renormalized dispersing band $G_f$. If the renormalized band is narrow enough (at, or close to, a partial Mott transition), the atomic features of the $G_d(\tau)$ can be resolved, whereas in the metallic regime usually $G_f$ is broad and those features are washed out after convolution. We use this method to show that various (renormalized) atomic multiplets can be identified in the complex many-body spectrum of the multi-channel Hubbard model. In addition to the bare atomic orbitals, this contains additional multiplets arising due to an interplay between Mott-localization in one band and itineracy in the other band.\,\cite{Nunez18} 
We demonstrate this explicitly, by comparing slave-particle methods to the solution of a two site problem considered before \cite{Nunez18} and then generalizing to the lattice. 

Inserting \pref{eq7} in \pref{eq6} and going to momentum space 
\bea
G_{d,\alpha\beta}(k,\tau)&=&Z_{\alpha\beta}G_{f,\alpha\beta}(k,\tau)\nonumber\\
&&\hspace{.5cm}+[\Pi_{\alpha\beta}(\tau)-Z_{\alpha\beta}]\sum_q G_{f,\alpha\beta}(q,\tau),\label{eq9}
\eea
i.e. $G_d$ contains a $k$-dependent part coming from renormalized non-interacting band $G_f$ plus some $k$-independent atomic structure in the last term. In particular in the Mott phase $Z=0$ no $k$-dependence exist.

\begin{figure}
\includegraphics[width=1\linewidth]{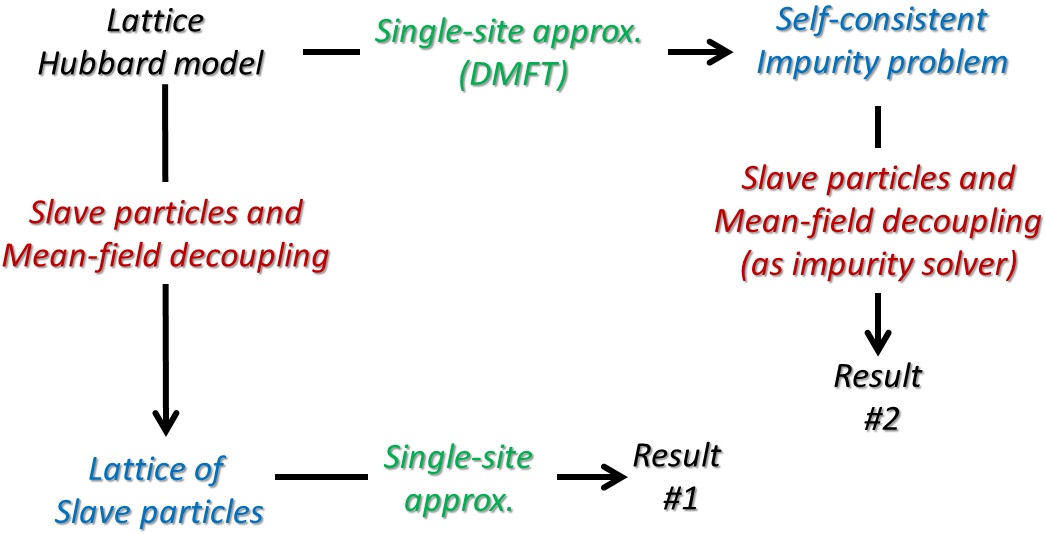}
\caption{A representation of non-commutativity of the two approximations: mean-field decoupling and single-site approximation.\small\label{fig:fig1}}
\end{figure}

A second point of this paper is that the two limits of mean-field decoupling and single-site approximation generally do not commute with each other. This is schematically represented in Fig.\,\pref{fig:fig1}.
To see this, consider doing a single-site approximation, first. This is achieved by the single-site DMFT, according to which the lattice problem is mapped to an effective impurity problem embedded within a conduction bath
\be
H_{imp}=\sum_{k\alpha\beta}(V_{k\alpha\beta}d\dg_\alpha c_{k\beta}+h.c.)+H_{int}+H_c,\label{eq9}
\ee
where $H_c=\sum_{k\alpha}\eps_{k\alpha}c\dg_{k\alpha}c\dn_{k\alpha}$ and the hybridization function $\Delta(z)=\sum_k|V_{k\alpha\beta}|^2/(z-\eps_k)$ is chosen so that locally, the Green's functions of the lattice and impurity are equal
\be
\sum_kG_{lat}(k,z)=G_d(z).\label{eq11}
\ee
The impurity problem is solved first. Extracting $\Sigma_I(z)$ from the $G_d^{-1}(z)=z-\Delta(z)-\Sigma_I(z)$, and assuming $\Sigma_{lat}(k,z)=\Sigma_I(z)$ is local, the lattice Green's function $G(k,z)=[z-\eps_k-\Sigma_I(z)]^{-1}$ is computed. From this and Eq.\,\pref{eq11}, a new hybridization function $\Delta(z)$ is extracted and the loops is repeated until convergence is reached. 

The slave-particle methods can be used as an impurity solver for this DMFT loop. The resulting Hamiltonian 
\be
H_{imp}=\sum_{k\alpha\beta}(V_{k\alpha\beta}\hat z\dg_\alpha f\dg_\alpha c_{k\beta}+h.c.)+H_{int}\{\hat z_{\alpha}\}+H_c\label{eq12}
\ee
is still interacting. %An NRG comparison between Eqs. \pref{eq12} and \pref{eq9} is provided in the next section. 
Various approximate schemes exist to solve this impurity problem, including non-crossing approximation (NCA)\,\cite{Costi96,Haule01} and one-crossing-approximation (OCA).\,\cite{Haule01} For the purpose of studying multiplets and the comparison to the mean-field solution above, it suffices to settle on a mean-field decoupling. As a result,   Eq.\,\pref{eq12} gives $H_{imp}=H_{imp,f}+H_{z,ss}$ where $H_{z,ss}$ is the same as in Eq.\,\pref{eq4}, but $H_{imp,f}$ is given by
\be
H_{imp,f}=\sum_{k\alpha\beta}(\tilde V_{k\alpha\beta}f\dg_\alpha c_{k\beta}+h.c.)+\sum_{k\alpha}\eps_{k\alpha}c\dg_{k\alpha}c\dn_{k\alpha}
\ee
where $\tilde V_{k\alpha\beta}=V_{k\alpha\beta}z^*_\alpha$. 
The impurity Green's function is then
\be
\hspace{-.2cm}G_{f,\alpha\beta}(z)=\frac{1}{z-\tilde\Delta_{\alpha\beta}(z)}, \quad G_{d,\alpha\beta}(\tau)=\Pi_{\alpha\beta}(\tau)G_{f,\alpha\beta}(\tau)
\ee
The mean-field parameters are obtained from minimization of $F=F_f+F_z(a)-2az$ and are given by\,\cite{Komijani17}
\be
\hspace{-.4cm}a_\alpha=-\frac{2}{z_\alpha}\int{\frac{d\omega}{\pi}}f(\omega)\im{G_f^{\alpha\alpha}(\omega+i\eta)}, \quad z_\alpha=\frac{dF_z}{da_\alpha}
\ee

It is clear that even though $G_d(\tau)$ is factorizable, in this scheme $G_{lat}(k,\tau)$ does not factorize, as opposed to \pref{eq6}, and consequently the multiplets are generally dispersing. Another manifestation of the non-commutativity of the two approximations is difference in $Z$  computed from the two approaches. We present a comparison of the two approaches for the single-orbital case in the next section.

It is noteworthy that under the commonly used simplification $\Pi(n,\tau)\approx Z$, then $\Sigma_I(z)=\Sigma_{lat}(z)=(1-Z)z$ and 
the two approximation schemes discussed above are equivalent as it can be shown explicitly.\,\cite{Komijani17} But the multiplets (the central focus of this paper) would be lost in this approximation.

%Mean-field single-site schemes have been used in the past to asses localization/itineracy of the orbitals and in particular to study the orbital selective Mott transitions with results that has been sometimes in contradiction to the DMFT. The use of DMFT with mean-field impurity solver allows us to revisit this problem in the last section of the paper.

In Kotliar-Ruckenstein (KR) four boson method,\,\cite{Kotliar86} or in rotationally invariant slave bosons (RISB),\,\cite{Lechermann07} the bosons are condensed (treated as $c$-numbers) at zero temperature, which is another (third) level of approximation, equivalent to $\Pi(n,\tau)\approx Z$ mentioned above. Since the time-dependence of $\Pi(\tau)$ is lost in this process, no atomic multiplet shows up in the spectra. Gaussian corrections to the condensate, simultaneously a) retrieves the Hilbert space of bosons, b) accounts for spatial and c) temporal fluctuations mentioned before.

%\section{NRG study of $Z_2$ slave-spin method}

\section{Z$_2$ Slave spin mean-field}
The above discussion was general. In this section, we focus on the Z$_2$ slave-spin method,\,\cite{Medici05} where at half-filling $z_\alpha=\tau^x_\alpha$ and the constraint $2f\dg_\alpha f\dn_\alpha=\tau^z_\alpha+1$ is applied on average, via a Lagrange multiplier $\lambda_\alpha$. Here $\tau_\alpha^a$ with $a=x,y,z$ are Pauli matrices that square to 1. Due to particle-hole symmetry, these Lagrange multipliers vanish $\lambda_\alpha=0$ at the saddle point.\,\cite{Komijani17}

The fact that the U(1) charge is carried by the $f_\alpha$ in this method, indicates that as long as $\tilde t_{ij}=t_{ij}\braket{\tau^x_i\tau^x_j}$ is non-zero (even if $\braket{\tau^x}=0$) the bulk is conducting and the system is not in a Mott phase.\,\cite{Nandkishore12} Therefore, we only consider the limit of large dimensions where single-site approximation is valid and the relation $\braket{\tau^x_i\tau^x_j}=\braket{\tau^x}^2$ is satisfied and $\braket{\tau^x}=0$ is equivalent to Mott localization.\,\cite{Zitko15,Komijani17}

\subsection{Single-band}
In the single-band case, the Hamiltonian has the general impurity form of Eq.\,\pref{eq9} with $\alpha=\ua,\da$ and the interaction $H_{int}=U\tilde n_\ua\tilde n_\da$ where $\tilde n_\sigma=d\dg_\sigma d\dn_\sigma-1/2$.  Using slave-spin we identify $\tilde n_\sigma=\tau^z_\sigma/2$, so that
\be
H_{z,ss}=a_\ua\tau^x_\ua+a_\da\tau^x_\da+U\tau^z_\ua\tau^z_\da
\ee
where $a_\sigma=2{\cal J}z_\sigma$ and ${\cal J}=-0.212 D$ is the average kinetic energy for a Bethe lattice of bandwidth $D=2t$. A better choice of basis is
\be
\hspace{-.5cm}\ket{\psi^{\pm}_1}=\frac{\ket{\Ua_\ua\Da_\da\pm\Da_\ua\Ua_\da}}{\sqrt 2},\quad \ket{\psi^\pm_2}=\frac{\ket{\Ua_\ua\Ua_\da\pm\Da_\ua\Da_\da}}{\sqrt 2},
\ee
{Here, $\ket{\Ua_\sigma}$ or $\ket{\Da_\sigma}$ refers to the eigen-states of $\tau^z_\sigma$ operator, where $\sigma=\ua,\da$.} In the paramagnetic regime $a_\ua=a_\da=a$ and with p-h symmetry, $\ket{\psi^-_{1,2}}$ decouples and in the basis of $\ket{\psi^+_{1,2}}$ the Hamiltonian reduces to $H_{z,ss}=\alpha\tau^x+(U/4)\tau^z$ with $\alpha=2a$. {This two-state problem can be diagonalized with an $SO(2)$ rotation
\be
\mat{\ket{\psi_g}\\\ket{\psi_e}}=\mat{\cos\theta & \sin\theta \\ -\sin\theta & \cos\theta}\mat{\ket{\psi_1^+} \\ \ket{\psi_2^+}}
\ee
where $\tan2\theta=2\alpha/(E_2-E_1)$} and the eigen-energies are
\be
E_{g/e}=U/4\mp\sqrt{(U/4)^2+\alpha^2}.\label{eq16}
\ee
Note that $E_g(\alpha\to0)\approx -2\alpha^2/U$. The mean-field study of this problem has been discussed in the past.\, \cite{Medici05,Medici17,Komijani17} In the case of the Hubbard model, {$z_\sigma=\braket{\psi_g\vert\tau^x_\sigma\vert\psi_g}=\sin 2\theta$. By minimizing the free energy, it can be shown that $Z\equiv \abs{z_\sigma}^2=\sin^22\theta$} follows the Brinkman-Rice theory \cite{Brinkman70} $Z=[1-u^2]\theta(1-u)$ where $\theta(x)$ is the Heaviside step function, $u=U/U_c$ and $ U_c=16\abs{\cal J}=3.36\times 2t$. In the case of an Anderson impurity, $Z=\abs{z_\sigma}^2$ plays the role of ``order parameter'' for the Kondo physics.\,\cite{Komijani17}
In both cases, the function $\Pi(\tau)$ at low temperature is equal to 
\bea
\Pi(\tau)&=&\braket{T\tau^x_\ua(\tau)\tau^x_\ua}=Z+(1-Z)e^{-\abs{\tau}\Delta E},\label{eq17}
\eea
where $\Delta E=E_e-E_g$ and $Z=\sin^2(2\theta_g)$. Multiplying by $G_f(\tau)$ and Fourier transforming
\bea
G_d''(\omega)&=&ZG''_f(\omega)+(1-Z)\Big[G''_f(\omega+\Delta E)\theta(\omega<-\Delta E)\nonumber\\
&&\hspace{1cm}+G''_f(\omega-\Delta E)\theta(\omega>\Delta E)\Big]\label{eq21}
\eea
Here, $\theta(\omega)$ appears as a low-temperature limit of $f(\omega\pm \Delta E)+n_B(\pm \Delta E)$ (Appendix C). The real part $G'_d(\omega)$ follows from Eq.\,\pref{eq21} using Hilbert transform.

\begin{figure}[tp!]
\includegraphics[width=\linewidth]
{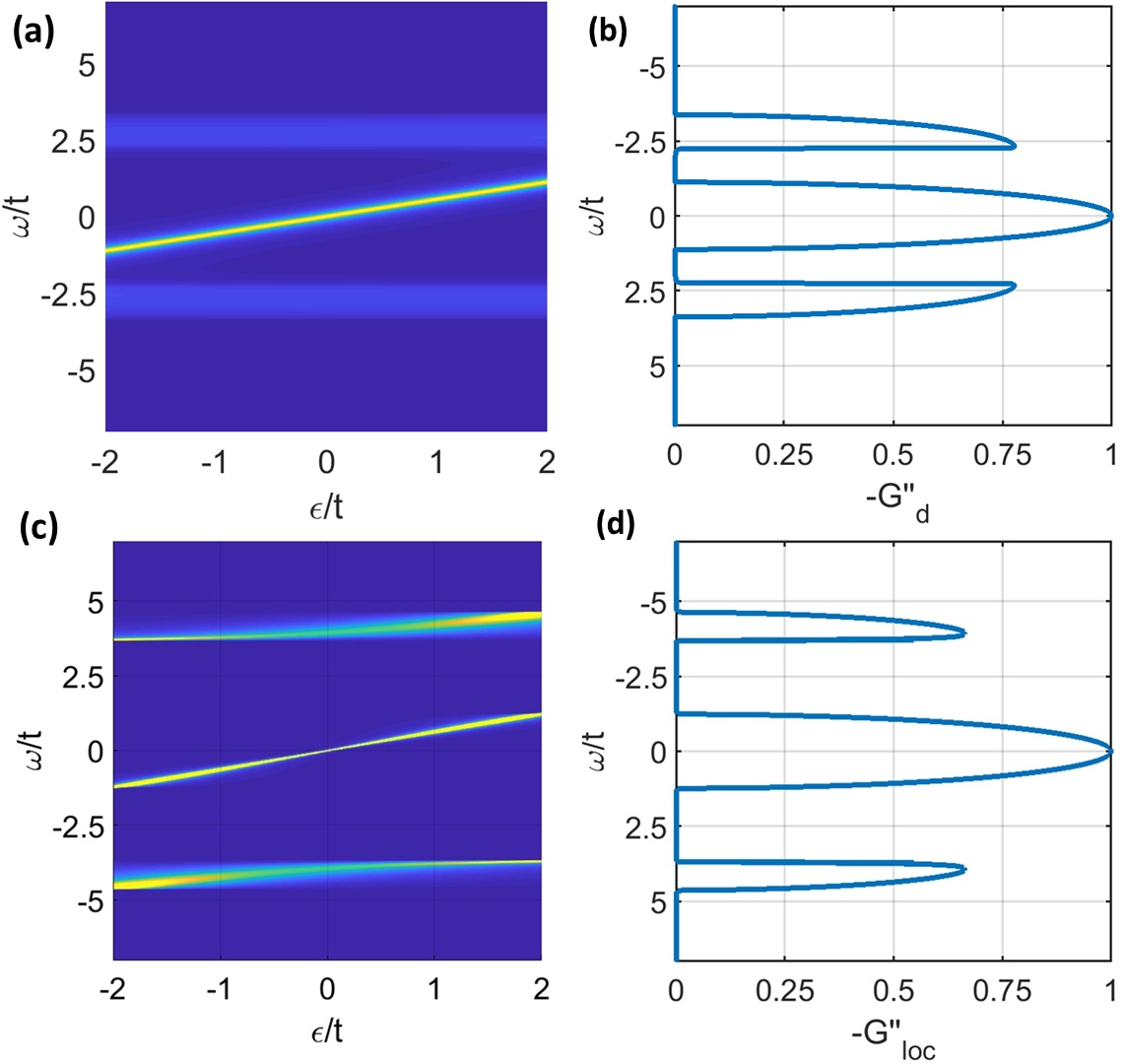}
\caption{\small A comparison of result \#1 and \#2 of Fig.\,\pref{fig:fig1} for a single-band system at $U=4.5t$. (a) The slave-spin mean-field spectral function $-G''(\epsilon,\omega+i\eta)$  as function of frequency and the Bethe lattice energy $\epsilon$. (b) The local spectral function, with $Z=0.56$.  (c) The result of using slave-spins as the impurity solver for DMFT which give $Z=0.63$. (d) The local spectral function from DMFT. }\label{fig:fig2}
\end{figure}

On a Bethe lattice of bandwidth $2t$ with density of states $\rho(\eps)=(\pi t)^{-1} \sqrt{1-(\epsilon/2t)^2}$, we can plot the spectrum as a function of frequency $\omega$ and the Bethe lattice energy $\epsilon$.
Fig.\,\pref{fig:fig2} shows a comparison between the ($\epsilon$-resolved and integrated) spectral function, as computed from Eq.\,\pref{eq21} with a Brinkman-Rice $Z$ and a lattice $G_f$, vs. a DMFT calculation with a slave-spin impurity solver where $G_f$ is the Greens' function of the impurity. Note that the Hubbard bands are featureless in (a) but disperse in (c) and there are slight differences in $Z$. However, close to the Mott transition, slave-spins behave poorly as the impurity solver for DMFT due to the fact that $\Delta(z)=z-\Sigma_I-G_{loc}^{-1}$ is non-analytical, the so-called {non-causality} of the impurity solver.

\subsection{Two-band model}
In the two-band model, the impurity problem is given by Eq.\,\pref{eq9} with $\alpha=1\ua,1\da,2\ua,2\da$. To make a connection to Ref.\,\cite{Nunez18} we choose the same form of simplified interaction 
\bea
H_{int}&=&U\sum_{i}\sum_{m=1}^2\tilde n_{im\ua}\tilde n_{im\da}+U_{12}\sum_{i}\sum_{\sigma\sigma'=\ua,\da}\tilde n_{i1\sigma}\tilde n_{i2\sigma'}\nonumber\\
&=&\sum_{i}\Big\{\frac{U}{2}(\tilde n_{i1\ua}+\tilde n_{i1\da}+\tilde n_{i2\ua}+\tilde n_{i2\da})^2-\frac{U}{2}\nonumber\\
&&\qquad\qquad\quad-\Delta(\tilde n_{i1\ua}+\tilde n_{i1\da})(\tilde n_{i2\ua}+\tilde n_{i2\da})\Big\}
\eea
where $\tilde n_\alpha \equiv n_{f\alpha}-1/2$ and  $\Delta=U-U_{12}$ is the difference between the intra and inter-orbital Coulomb couplings. We will drop the $-U/2$ constant term in the following. While this form of the interaction is simpler to follow, we have obtained qualitatively similar results with a more general Kanamori Hamiltonian. Within slave-spin method this becomes
\be
H_{z,ss}=\sum_{m\sigma}a_{m\sigma}\tau^x_{m\sigma}+H_{int}[\tilde n_\alpha\to\tau^z_\alpha]\label{eq24}
\ee
As before, $a_{m\sigma}=2{\cal J}_mz_{m\sigma}$ and ${\cal J}_m\propto D_m$ and in the paramagnetic regime, $a_{m\sigma}=a_m$ and $z_{m\sigma}=z_m$. The Hamiltonian is a $16\times16$ matrix, 
and the full calculation of the wavefunctions, eigen-energies and the correlation functions $\Pi_{11}(\tau)$ and $\Pi_{12}(\tau)$ are done numerically.

\begin{figure}[tp!]
\includegraphics[width=0.7\linewidth]{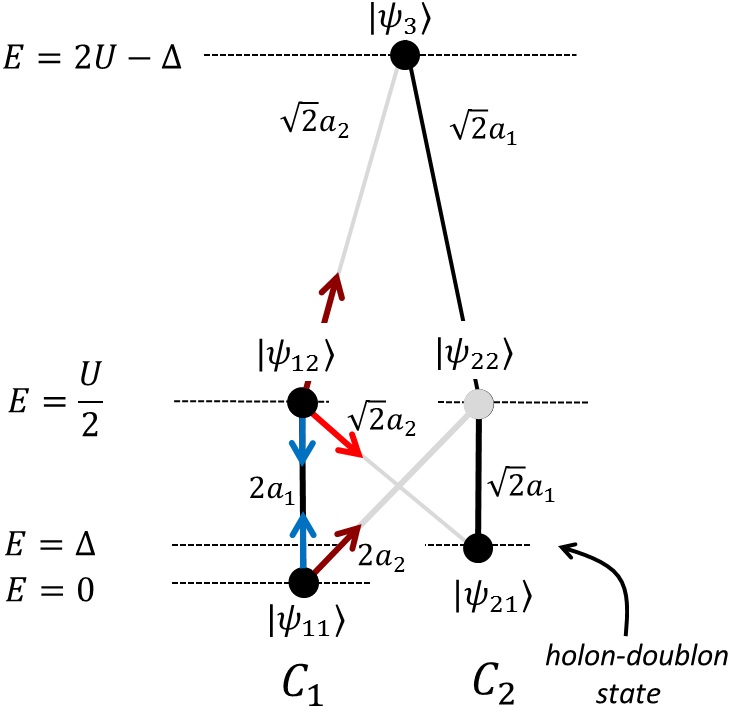}
\caption{\small The relevant sector of the slave-spin Hamiltonian (see Appendix B), with states given in Eqs.\,\pref{eq25}. At OSM phase $a_2=0$, and the Hamiltonian breaks into $C_1$ and $C_2$ sectors, with the ground state in sector $C_1$ due to larger level-repulsion and lower energies. The transitions caused by $\tau^x_{1\sigma}$ and $\tau^x_{2\sigma}$ are indicated by blue and red arrows, respectively. The bright-red arrow is to the holon-doublon bound-state $\ket{\psi_{21}}$.
}\label{fig:fig3}
\end{figure}

Here in order to get an analytical insight, we rather make a simplifying assumptions that the system is already in an OSM regime, $a_2=2{\cal J}_2z_{2}=0$. As a result, the relevant sectors of the Hamiltonian (Appendix B), shown diagrammatically in Fig.\,\pref{fig:fig3} breaks into two $2\times 2$ matrix blocks with level repulsion $\alpha_1=2a_1$ and $\alpha_2=\sqrt{2}a_1$. 
The vertical axis is the energy. The diagonal elements of the matrix are represented by black/gray dots and the off-diagonal matrix elements are represented by the black/gray lines connecting the dots. The basis and the diagonal energies are
\bea
\ket{\psi_{11}}&=&\frac{\ket{\Ua_{\ua}\Da_{\da}+\Da_\ua\Ua_\da}_1}{\sqrt 2}\frac{\ket{\Ua_\ua\Da_\da+\Da_\ua\Ua_\da}_2}{\sqrt 2},\quad E_{11}=0\nonumber\\
\ket{\psi_{21}}&=&\frac{\ket{\Ua_\ua\Ua_\da}_1\ket{\Da_\ua\Da_\da}_2+\ket{\Da_\ua\Da_\da}_1\ket{\Ua_\ua\Ua_\da}_2}{\sqrt 2},\quad E_{21}=\Delta\nonumber\\
\ket{\psi_{12}}&=&\frac{\ket{\Ua_\ua\Ua_\da+\Da_\ua\Da_\da}_1}{\sqrt 2}\frac{\ket{\Ua_\ua\Da_\da+\Da_\ua\Ua_\da}_2}{\sqrt 2}, \quad E_{12}=\frac U2\nonumber\\
\ket{\psi_{22}}&=&\frac{\ket{\Ua_\ua\Da_\da+\Da_\ua\Ua_\da}_1}{\sqrt 2}\frac{\ket{\Ua_\ua\Ua_\da+\Da_\ua\Da_\da}_2}{\sqrt 2},\quad E_{22}=\frac U2\nonumber\\
\ket{\psi_{3}}&=&\frac{\ket{\Ua_\ua\Ua_\da}_1\ket{\Ua_\ua\Ua_\da}_2+\ket{\Da_\ua\Da_\da}_1\ket{\Da_\ua\Da_\da}_2}{\sqrt{2}},\quad E_3=2U\qquad
\label{eq25}
\eea
In the following we assume that the fully empty/filled states $\ket{\psi_3}$ can be discarded. This is justified close to the Mott transition.\,\cite{Komijani17} In terms of these, the eigen-functions $j=1,2$ for the sector $i=1,2$ are
{\be
\mat{\vert{\tilde\psi_{i1}}\rangle\\\vert{\tilde\psi_{i2}}\rangle}=\mat{\cos\theta_i & \sin\theta_i \\ -\sin\theta_i & \cos\theta_i}\mat{\vert{\psi_{i1}\rangle} \\ \vert{\psi_{i2}}\rangle}
\ee
where $\tan2\theta_{i}=2\alpha_i/(E_{2i}-E_{1i})$} and the eigen-energies
\be
\tilde E_{ij}=\frac{E_{i1}+E_{i2}}{2}-\sigma_j\sqrt{[(E_{i1}-E_{i2})/2]^2+\alpha_i^2}\label{eq26}
\ee
where $\sigma_1=-\sigma_2=1$. The level-repulsions are $\alpha_1=2a_1$ and $\alpha_2=\sqrt{2}a_1$. The ground state belongs to the sector $i=1$. The correlation functions $\Pi_{ii}(\tau)\equiv\langle T\tau^x_{i\ua}(\tau)\tau^x_{i\ua}\rangle$ are then
\bea
\Pi_{11}(\tau)&=&Z_1+(1-Z_1)e^{-\abs{\tau}(\tilde E_{12}-\tilde E_{11})}\label{eq27}\\
\Pi_{22}(\tau)&=&e^{-\abs{\tau}(\tilde E_{21}-\tilde E_{11})}\sin^2(\theta_{2}+\theta_{1})\nonumber\\
&&\qqquad+e^{-\abs{\tau}(\tilde E_{22}-\tilde E_{11})}\cos^2(\theta_{1}+\theta_{2}).\nonumber
\eea
where as before $Z_1=\sin^2(2\theta_{1})$.
\subsubsection*{Holon-doublon peaks in orbital selective Mott phases}
Eqs.\,\pref{eq25} and \pref{eq27} show that the lowest-energy intermediate state accessed by $\Pi_{22}$ is the doublon-holon state $\ket{\psi_{21}}$. %Whereas the other states are product states of two different orbitals, the doublon-holon state is an entangled state of a doublon occupancy in either first or second orbital and a holon in the other one. 

\begin{figure}[tp!]
\includegraphics[width=1\linewidth]{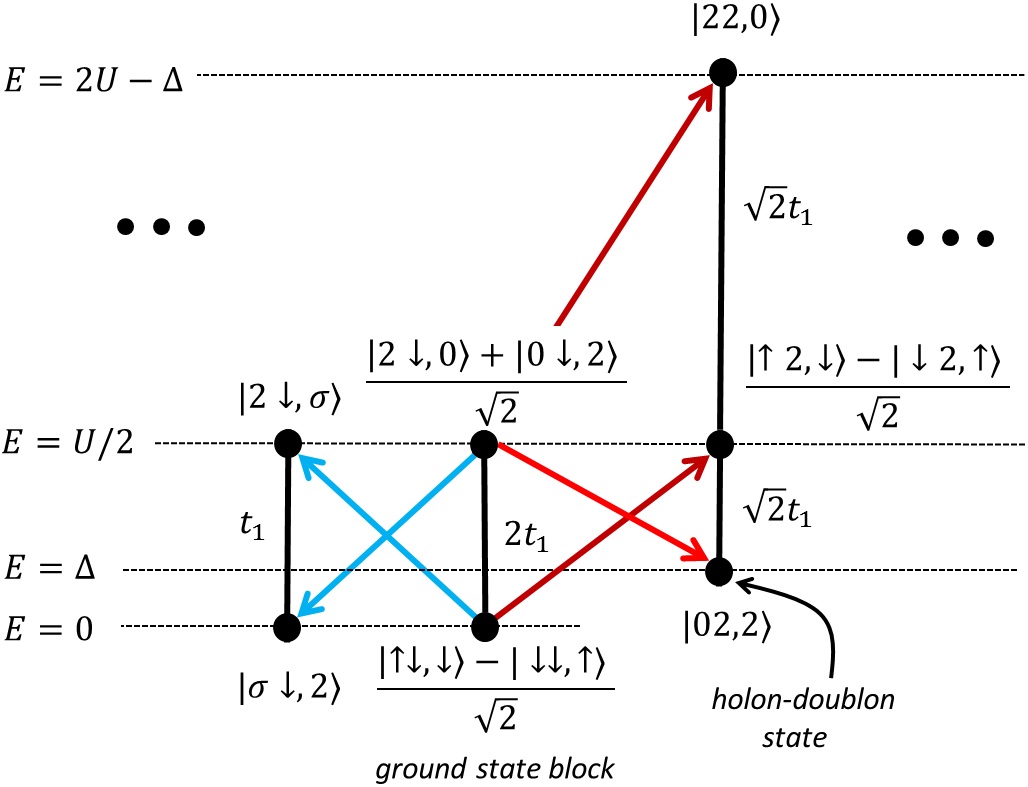}
\caption{\small Exact diagonalization of a two-site problem. The relevant sectors of the Hamiltonian of a two-orbital impurity with one bath site (Appendix A) when $t_2=0$. The states of the impurity $\alpha_1$ and $\alpha_2$ and the bath $\beta_1$ are indicated in the form $\ket{\alpha_1\alpha_2,\beta_1}$. Blue and red arrows indicate the transitions caused by $d\dg_{1\sigma}$ and $d\dg_{2\sigma}$, respectively. The bright red transition is to the holon-doublon state {at energy $\Delta$}, only accessible from the ground state if $t_1$ is non-zero. The ellipses on the left and right indicate other sectors of the Hamiltonian, not involved in the ground state and not accessible with $d\dg_{1,2}$.
}\label{fig:fig4}
\end{figure}

{This state is only accessible to the spectral function via the excited state $\ket{\psi_{12}}$, i.e. when the first  (wider) band is metallic.} It is instructive to study this effect on a two-site problem where one site is interacting and the other site plays the role of a non-interacting bath.\,\cite{Nunez18} This is studied explicitly in the Appendix A, and here we discuss the main result. Fig.\,\pref{fig:fig4} shows the relevant sectors of a two-site Hamiltonian when $t_2=0$ (emulating Mott localization of the second band). The blue and red arrows are the transitions caused by $d\dg_{1\sigma}$ and $d\dg_{2\sigma}$, respectively. The ground state is in the central sector and contains an admixture of the excited state due to the off-diagoanl mixing $2t_1$. The spectral function of the first (delocalized) orbital, probed by the blue arrows, contains renormalized Hubbard peaks as well as a peak near zero frequency which would evolve into an Abrikosov-Suhl resonance when the number of bath sites increases.\,\cite{Hewson} The holon-doublon state $\ket{02,2}$, is only accessible via the excited state admixture (shown in bright red) and would disappear for $t_1=0$. In other words, in the fully atomic limit, $\ket{02,2}$ is a dark state and is only visible when $t_1\neq 0$.

The slave-spin method is capable of capturing the correlations discussed above in the two-site problem (Appendix A) as well as the DMFT results of N\'u\~nez Fern\'andez et al.\,\cite{Nunez18} for the Hubbard model. Within single-site approximation, the Hamiltonian of the slave-spin sector is the same in all these cases (Eq.\,\ref{eq24} and Fig.\,\ref{fig:fig3}), while the mean-values of the parameters $a_1$ and $a_2$ are different. The transitions probed by the the $\tau^x_{2\sigma}$ in the slave-spin Hilbert space are marked by red in Fig.\,\pref{fig:fig2}. The action of $\tau^x_{2\sigma}$ on $\ket{\psi_{11}}$ leads to $\ket{\psi_{21}}$ whereas its acting on $\ket{\psi_{12}}$ leads to $\ket{\psi_{22}}$ (bright red). Therefore, only if the ground state contains admixture of $\ket{\psi_{12}}$, i.e. when $\theta_1\neq 0$ and the first band is metallic, do the holon-doublon state appear in the spectral function of the second orbital, or vice versa.

Assuming that the second orbital is in the Mott phase, in the impurity model $G_{f2}(\tau)=-\sgn\tau/2$, and in the lattice model $G_{f2}(n,\tau)=-\delta_{n0}\sgn\tau/2$. Multiplying by $\Pi_{22}(\tau)$ and Fourier transforming we find
\be
2G_{d2}({\rm z})=\frac{\sin^2(\theta_{2}+\theta_{1})}{{\rm z}-(\tilde E_{21}-\tilde E_{11})}+\frac{\cos^2(\theta_{1}+\theta_{2})}{{\rm z}-(\tilde E_{22}-\tilde E_{11})}-({\rm z}\to-{\rm z}).\label{eq28}
\ee
The first term is the doublon-holon peak observed by the DMFT\,\cite{Nunez18} in the spectrum of the narrow band in OSM regime. The spectrum of the wider (itinerant) band is not affected and is essentially given by the results of previous section, Eq.\,\pref{eq21}, up to an enhancement of the effective Coulomb energy $U/2\to U/2+\Delta$ by the inter-orbital interaction. 

%Fig.\,\pref{fig5} shows the spectrum of the two bands $-G''_{d1}(\omega)$ and $-G_{d2}''(\omega)$ for $t_2/t_1=0.1$ at $U=0.74U_{c1}$, as $\Delta/U_{c1}$ is varied. Here, $U_{c1}=16\abs{{\cal J}_1}=6.79t_1$. According to slave-spin mean-field theory \cite{Medici05,Komijani17} the system is in the OSMP regime.  The Hubbard peaks at higher frequencies and the doublon-holon peaks at lower frequencies are visible. 

\begin{figure}
\includegraphics[width=1\linewidth]{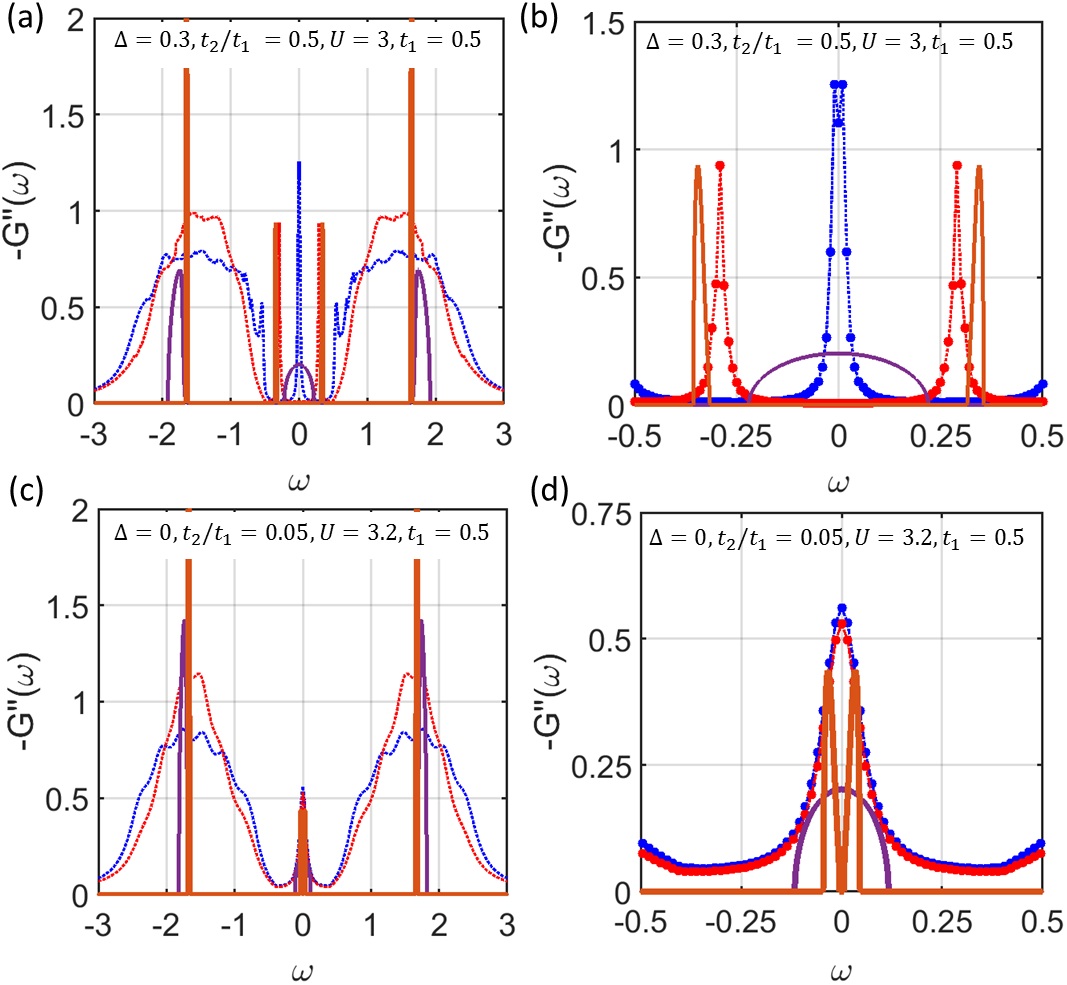}
\caption{\small A comparison of the spectral results between slave-spin mean-field (continuous lines in dark color) and DMFT+DMRG (dashed lines without/with data points in left/right panels in bright color). The spectrum of wider/narrower channels are shown in blue/red color. (a) In presence inter-orbital interaction $\Delta=0.3$, $t_2/t_1=0.5$ $U=3$ and $t_1=0.5$. 	(b) A zoom into low frequency part of (a). (c) No inter-orbital interaction $\Delta=0$, $t_2/t_1=0.05$, $U=3.2$ and $t_1=0.5$. (d) A zoom into low frequency part of (c).} 
 \label{fig5}
\end{figure}

\subsubsection*{Comparison with DMFT and discussion}

Fig.\,\pref{fig5} compares the result of calculation of spectrum between slave-spin with that of DMRG+DMFT.\, \cite{Nunez18} Fig.\,\pref{fig5}(a) shows a comparison of the spectra in the OSMP regime in presence of inter-orbital interaction $\Delta=0.3$ and $t_2/t_1=0.5$. The incoherent broadening of the Hubbard peaks are not captured in the mean-field theory. However, a zoom into the low-frequency part in Fig.\,\pref{fig5}(b) shows that there is a good agreement on the position and the amplitude of the holon-doublon resonance between the two methods. 

In spite of the qualitative agreements for $\Delta\neq 0$,  there are some disagreements for $\Delta=0$ between slave-spin mean-field predictions and the DMFT+DMRG numerics. Fig.\,\pref{fig5}(c) compares the spectra for $\Delta=0$ and large anisotropy $t_2/t_1=0.05$. The slave-spin mean-field predicts OSMP in this regime, whereas DMFT predicts a finite $Z$. A zoom into low-frequency part of spectra in Fig.\,\pref{fig5}(d) shows that in contrast to DMFT result, for $\Delta=0$ the spectrum of the narrow band remains gapped and the wavefunction renormalization remains zero [also Fig.\,\pref{fig5}(a)] in the mean-field solution. 

The origin of this gap can be traced back to the pole ${\rm z}=\tilde E_{21}-\tilde E_{11}$ in Eq.\,\pref{eq28}. When $\Delta\to 0$, this gives
\be
{\rm z}\to\sqrt{(U/4)^2+4a_1^2}-\sqrt{(U/4)^2+2a_1^2}.
\ee
{If $a_1\gg U$ this is at ${\rm z}=(2-\sqrt 2)a_1$ and for $a_1\ll U $ is  ${\rm z}\to 4a_1^2/U$, which linearly or quadratically depends on the width of the wider channel (or $T_{K1}$ in the case of impurity). Therefore, the peak is expected to remain at finite frequency and follow the width of the coherent band in the wider channel.

Equations \,\pref{eq27} and \pref{eq28} show that although the total spectral weight of the two orbitals is equal to one, $\sin(2\theta_{1})\neq\sin(\theta_{1}+\theta_{2})$ and thus the weight of two holon-doublon peaks in the second orbital is not equal to the weight of the coherence band in the first orbital, in contrast to the observation of Ref.\,\cite{Nunez18}. $\theta_i\in(0,\pi/4)$ quantifies the admixture of high-energy state in the ground state of block $i$. For $\Delta/U\ll 1$, we have $\theta_2<\theta_1$ and the coherence peak has higher spectral weight than the doublon-holon peaks. But for $2\Delta/U>1-1/\sqrt 2$ this trend reverses.}

{When both bands are metallic, all five states mix to create various eigen-states. When $a_2$ is small, we can assume that the energies of $\vert\tilde\psi_{ij}\rangle$ are only slightly modified from OSMP regime and $\vert\tilde\psi_{11}\rangle$ is still the ground state. However, the low-lying excited state $\vert\tilde\psi_{21}\rangle$ receives some admixture of $\ket{\psi_{12}}$ of $O(a_2)$. Therefore, $\langle\tilde\psi_{21}\vert \tau^x_{1\sigma}\vert\tilde\psi_{11}\rangle=O(a_2)$ is non-zero and therefore, a weak resonance feature appears at energy $\tilde E_{21}$ in the function $\Pi_1(z)$. This feature is further weakened due to convolution with the coherence band $G_f(z)$ and appears as slight modulation of the coherence band in the wider orbital.

\subsubsection*{Further numerical results from Slave-spin}
In this section, we summarize the numerical solution to the slave-spin mean-field equations. Fig.\,\pref{fig6} shows the evolution of quasiparticle peaks with Hubbard $U$ for $\Delta=0$, equivalent to zero Hund's coupling in the Kanamori-Hubbard model. Figs.\,\ref{fig6}(a,b) show the case of $t_2/t_1=0.1$ which contains the OSMP regime (according to mean-field). The wide band spectrum shows the coherent peak as well as renormalized Hubbard peaks. In the Narrow band spectrum the coherent peak disappears at $U/U_{c1}\sim0.2$ (top inset) while a doublon/holon resonant feature appears at $\omega/t_1\sim 0.5$ which follows the evolution of the coherent band in the first (wider) orbital, and going to zero when the first orbital enters the Mott phase. Figs.\,\ref{fig6}(c,d) show the case of $t_2/t_1=0.3$ which is the locking regime (bottom inset). The spectrum of the wide band is similar to before, but the narrow band is different in that a) There is a coherent peak at $\omega\sim 0$. Instead of doublon-holon peak, we have a doublon-holon band whose splitting follows the width of coherent band in the first orbital. There are additional fine structures in the Hubbard peaks in this case.

\begin{figure}[h!]
\includegraphics[width=\linewidth]{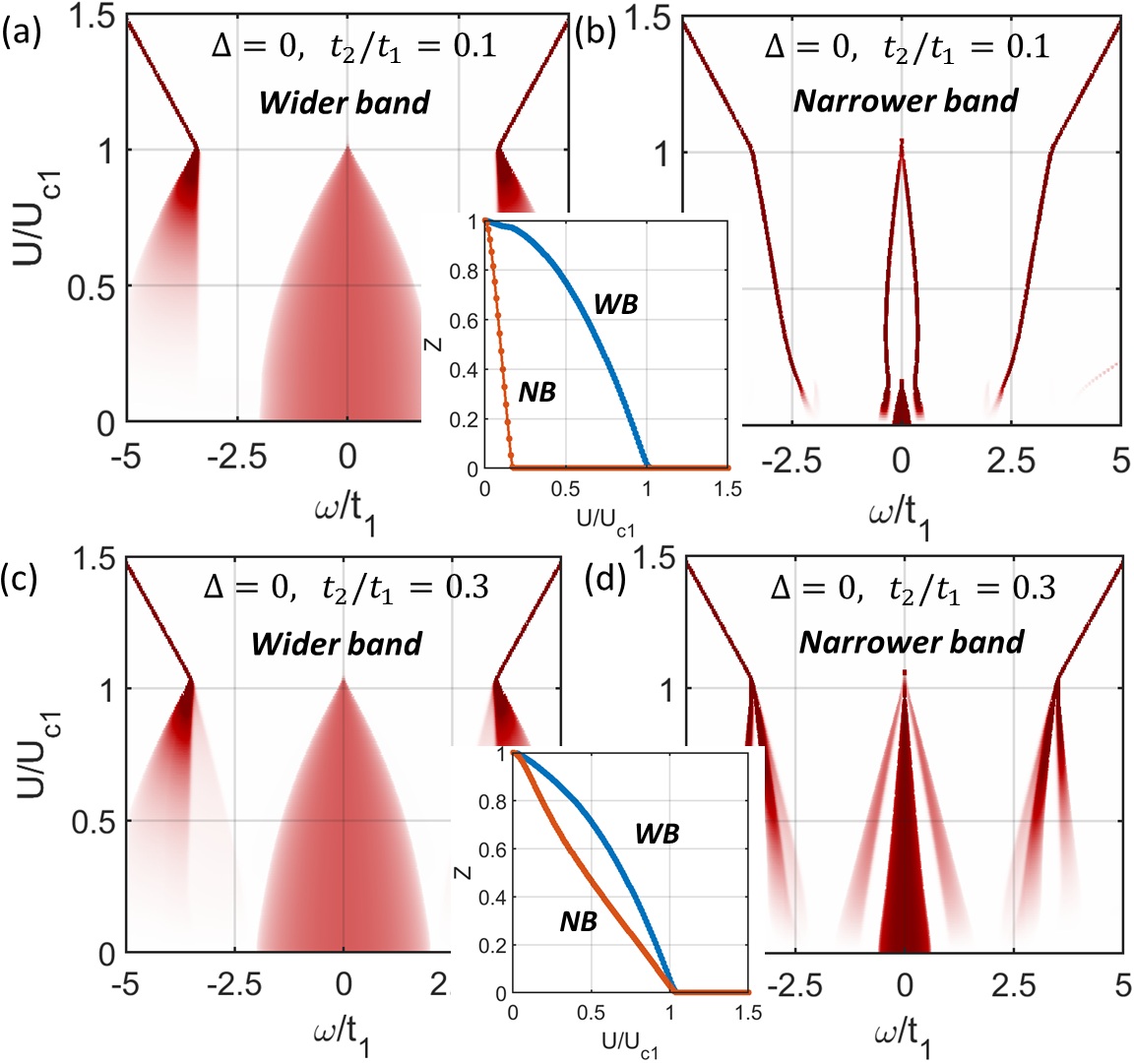}
\caption{\small The spectrum of the two bands in absence of inter-orbital interaction $\Delta=0$. (a,b) OSMP driven by bandwidth anisotropy and $t_2/t_1=0.1$. The middle inset shows the wavefunction renormalizations vs. $U/U_{c1}$ for the first (blue) and second (red) bands. (c,d) The locking regime $t_2/t_1=0.3$. The doublon-holon peaks develop into {\it  bands} in the spectrum of the narrower band in addition to the coherence band. Middle inset: wavefunction renormalizations. }\label{fig6}
\end{figure}

Figs.\,\ref{fig7}(a,b) show the evolution of quasi-particle peaks in the narrower (second) band in presence of the inter-orbital interaction $\Delta/U_{c1}=0.05$ for (a) $t_2/t_1=0.1$ and (b) $t_2/t_1=0.3$ with the wavefunction renormalizations shown in the insets. The spectrum of the wider band is similar to the $\Delta=0$ case. In both cases the doublon/holon quasiparticles are present but they disappear (at the Mott transition of the wider band) while their splitting is still finite.

%Fig.\,\ref{fig7}(c) shows the effect of breaking p-h symmetry by application of a chemical potential at $t_2/t_1=0.3$, $\Delta/U_{c1}=0.05$ and $U/U_{c1}=0.7$ (the dashed line in Fig.\,\ref{fig7}b). This is achieved~\cite{Hassan10} by a modification  $\tau^x_\alpha \to(\tau^+_\alpha+c\tau^-_\alpha)/2$ with the constant $c$ determined self-consistency, and including a Lagrange multiplier to enforce $2f\dg_\alpha f\dn_\alpha-1=\tau^z_\alpha$ in average (see Appendix D). We see that particle/hole-doping lifts the (orbitally selective) Mott transition and results in non-zero $Z_2$, a general property of slave-spin mean-field solutions.\,\cite{Hassan10} Fig.\,\ref{fig7}(d) shows the effect of a chemical potential on the spectrum of the narrower band. As a result of breaking p-h symmetry, a coherent band appears at $\omega\sim0$ and the holon/doublon resonance (only positive frequencies shown) splits, and broadens into two bands. \blue{This is in contrast to the Mott phase, where there is a threshold on $\mu~O(U)$, below which the density remains at half-filling. In the OSMP regime, the mean-field solution suggests that applying slightest $\mu$, lifts the OSMP.}

%The splitting of the doublon-holon peak can be roughly understood within the slave-spin Hamiltonian discussed in Appendix B. In the OSMP regime and with p-h symmetry, the slave-spin Hamiltonian decouples into two degenerate copies $H_z=H_z^+\oplus H_z^-$. A chemical potential couples the two copies and lifts the degeneracies, resulting in a peak-splitting.

\begin{figure}[tp!]
\includegraphics[width=\linewidth]{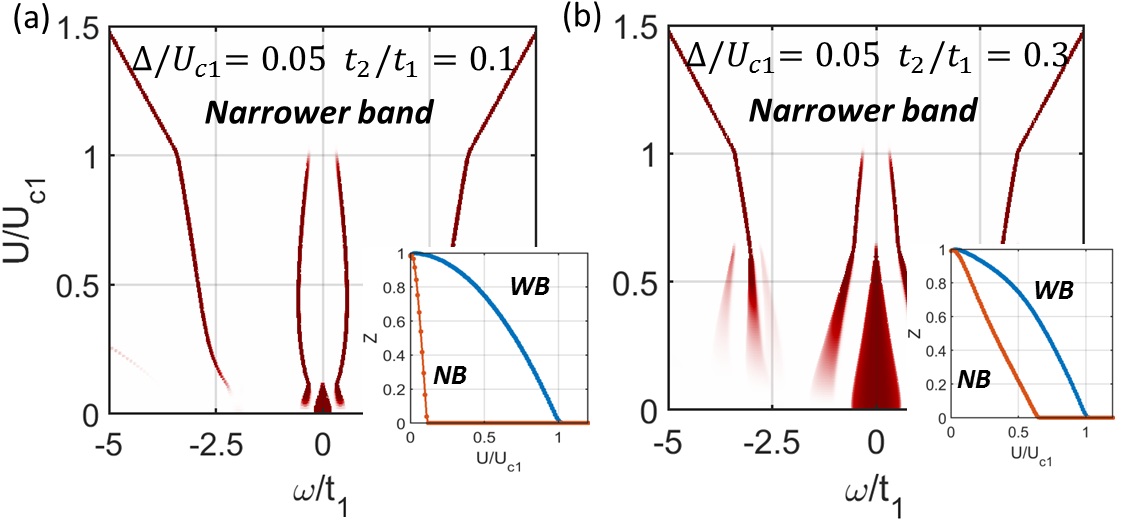}
\caption{\small 
The spectrum of the narrower band in presence of inter-orbital interaction $\Delta=0.05U_{c1}$: (a) $t_2/t_1=0.1$ and (b) $t_2/t_1=0.3$.  OSMP driven by bandwidth anisotropy and $\Delta$. The insets shows the wavefunction renormalizations vs. $U/U_{c1}$. 
%(c) The effect of breaking $p-h$ asymmetry by applying a chemical potential on $Z_1$ and $Z_2$ at $U=0.7U_{c1}$ and $\Delta=0.05U_{c1}=0.3 t_1$ indicated by the dashed line in (b). (d) The spectrum of narrow band evolves with chemical potential at $U=0.7U_{c1}$ and $\Delta=0.05U_{c1}=0.3t_1$: A coherence peak appears and the doublon-holon peak splits.
}\label{fig7}
\end{figure}

\section{Conclusions}
We have introduced a general formalism, by which, slave particle methods capture atomic multiplets in the spectrum of nearly Mott localized orbitals. We studied the commutativity of the mean-field decoupling and single-site approximation, showing that the multiplets get renormalized and acquire dispersion within DMFT. We used slave-spins and applied our formalism to reproduce the holon-doublon peak found in the DMFT results of N\'u\~nez Fern\'andez et al.\,\cite{Nunez18} for the two-band Hubbard model. Overall there is a good agreement between DMFT and slave-spin mean-field. However, in contrast to DMFT, the splitting between holon-doublon peaks in slave-spin mean-field solution does not go to zero in the limit of zero Hund's coupling, in consistency with an OSM phase. 
This raises the question of whether quantum fluctuations beyond mean-field can destroy the OSMP in absence of Hund's coupling, which we leave for the future.

The authors are grateful to Y.~N\'u\~nez Fern\'andez for fruitful discussions, providing data and an earlier collaboration, and also acknowledge stimulating discussions with P.-Y.~Chang and T.-H.~Lee. This work used XSEDE, which is supported by NSF grant ACI-1548562. K.~H acknowledges support from grants PICT 2016-0402 from the Argentine ANPCyT
and PIP 2015-2017 (CONICET). G.~K was supported by NSF grant DMR-1733071 and Y.~K was supported by a Rutgers University Materials Theory postdoctoral fellowship.
\section*{Appendix A - Two site problem}
\subsection*{1. Single-orbital case}
The Hamiltonian is $H=H_0+H_T+H_c$ where
\be
H_0=\frac{U}{2}(n_\ua+n_\da)^2, \qquad H_T=\sum_{\sigma}(t d\dg_\sigma c\dn_\sigma+h.c.)
\ee
In the two site problem, at half-filling we have $H_c=0$. The Hamiltonian has $SU_{\rm charge}(2)\otimes SU_{\rm spin}(2)$ symmetry. Anticipating future symmetry-lowering additions, we use a smaller $U_{\rm charge}(1)\otimes U_{\rm spin}(1)$ symmetry to label states with $Q^z$ and $S^z$. The distinct atomic states are denoted by filled circles at corresponding energy in Fig.\,\pref{fig:fig6}, and the transition between them by $H_T$ are marked in black. As a result, the Hamiltonian is block diagonal and the largest block is $2\times 2$ corresponding to three two-level systems on the right of Fig.\,\pref{fig:fig6}. Each group of atomic states connected by lines form a block. The larger the off-diagonal element of the block, the larger is the level-repulsion. Therefore, the ground state is given by the rightmost block:
\be
E_g=U/4-\sqrt{(U/4)^2+4t_1^2}
\ee
\be
\ket{\psi_g}=\cos\theta_g\frac{\ket{\ua,\da}+\ket{\da,\ua}}{\sqrt 2}+\sin\theta_g\frac{\ket{2,0}+\ket{0,2}}{\sqrt 2}\label{eqpsig}
\ee
with $\theta_g=\tan^{-1}(E_g/2t_1)$. The Green's function is
\be
G(\tau)=\braket{-Td_\ua(\tau)d\dg_{\ua}(0)}
\ee
At zero temperature and positive time $\tau>0$ we have
\be
G_D(\tau)\equiv G(\tau>0)=-\braket{\psi_g\Big\vert e^{\tau H}d\dn_\ua e^{-\tau H}{\mathbb 1} d\dg_\ua\Big\vert \psi_g}
\ee
The blue lines in Fig.\,\pref{fig:fig6}
show the transition caused by acting with $d\dg_1$ on the ground state block. The result is creation of a `doublon' at the impurity site, which belongs to the second rightmost block. The intermediate states in $\bb 1$ are
\be
\ket{\psi_\pm}=\cos\theta_\pm\ket{\sigma,2}+\sin\theta\ket{2,\sigma}
\ee
with the energies
\be
E_\pm=U/4\pm\sqrt{(U/4)^2+t_1^2}
\ee
where $\theta_\pm=\tan^{-1}(E_\pm/t_1)$.
Therefore, we find
\be
G_d({\rm z})=\frac{1}{2}\sum_{a=\pm}\Big[\frac{\sin^2(\theta_g+\theta_{\pm})}{{\rm z}-(E_a-E_g)}-({\rm z}\to -{\rm z})\Big],\label{eq25b}
\ee
where we used that the matrix element is given by $\langle{\psi_a\vert d\dg_\ua\vert \psi_g}\rangle=\sin(\theta_a+\theta_g)/\sqrt{2}$. 
%It can be seen that for $\tau<0$ the Green's function involves dynamics of a holon in the third rightmost block, and $G_H({\rm z})=-G_D(-{\rm z})$ and together, we have $G({\rm z})=G_D({\rm z})-G_D(-{\rm z})$. 
%\be
%G({\rm z})=\sum_{a=\pm}\Big[\frac{\abs{m_a}^2}{{\rm z}-(E_a-E_g)}+\frac{\abs{m_a}^2}{z+(E_a-E_g)}\Big]
%\ee
This spectrum is composed of two resonances symmetric w.r.t. $\omega=0$. The two energies have simple approximations in the limit of large $U$: one at $\sim t^2/U$ and the other at $\sim U/2$. The low-frequency resonance at $\omega=E_--E_g$ is the single-bath site signature of Abrikosov resonance peak (metallic regime), whereas the high-frequency resonance at $\omega=E_+-E_g$ is the renormalized Hubbard peak. 

\subsubsection*{Slave-spin}
It is remarkable that the same structure comes from slave-spin method. After mean-field decoupling of slave-spins and the quasiparticles we find $H=H_f+H_{z,ss}-2az$ where $H_f=\sum_\sigma(\tilde tf\dg_\sigma c_\sigma+h.c.)$ and $\tilde t_\sigma=z_\sigma t$. Diagonalizing $H_f$ using molecular bonding/anti-bonding states we find
\be
H=\tilde t (f_{+\sigma}\dg f\dn_{+\sigma}-f_{-\sigma}\dg f\dn_{-\sigma}),\quad \sqrt{2}f_\pm=f\pm c
\ee
The mean-field parameter $z$ can be worked out from minimizing $F(a,z)=F_f+E_S(a_\sigma)-\Sigma_\sigma a_\sigma z_\sigma$. Eliminating $a$ gives \cite{Komijani17}
\be
E_S=\frac{U}{4}[1-\frac{1}{\sqrt{1-z^2}}]
\ee
and
\be
F(z)=F_f(z)-\frac{U}{4}[1-\sqrt{1-z^2}]
\ee
In the present problem $F_f=-2_stz$. Therefore, 
\be
z=[1+(U/8t)^2]^{-1/2}
\ee
The Green's function $G_f(\tau)=\braket{-Tf(\tau)f\dg}$ for $\tau>0$ is
\be
G_f(\tau)=-\frac{1}{2}\sgn\tau e^{-\tilde t\abs{\tau}}
\ee
Multiplying this by $\Pi(\tau)$ from Eq.\,\pref{eq17} we find
\be
G_d(z)=\frac{1}{2}\Big[\frac{Z}{{\rm z}-\tilde t}+\frac{1-Z}{{\rm z}-(\tilde t+\Delta E)}-({\rm z}\to-{\rm z})\Big]\nonumber
\ee
which again has the two peak structure we saw previously.

\begin{figure}[h!]
\includegraphics[width=1\linewidth]{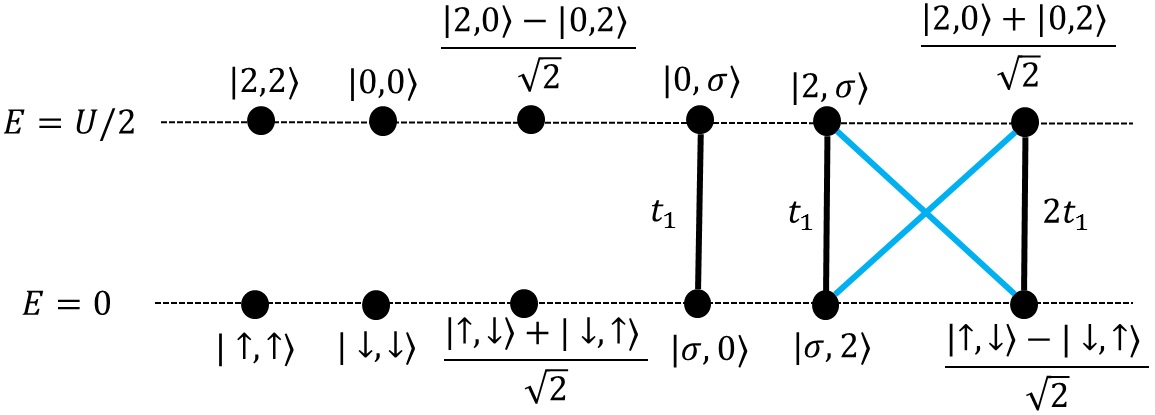}
\caption{\small\raggedright A representation of the Hamiltonian in the single-orbital two-site problem. $\ket{\alpha,\beta}=\ket{\alpha}_d\ket{\beta}_c$ are the atomic states of the dot $d$ and conduction site $c$, where $\alpha,\beta=0,\ua,\da,2$. Each atomic state is marked by a filled circle at the corresponding atomic energy. The black line denotes the transitions caused by tunnelling, whereas the blue lines, are transitions probed in the Green's function.\label{fig:fig6}}
\end{figure}
\begin{figure}[h!]
\includegraphics[width=0.5\linewidth]{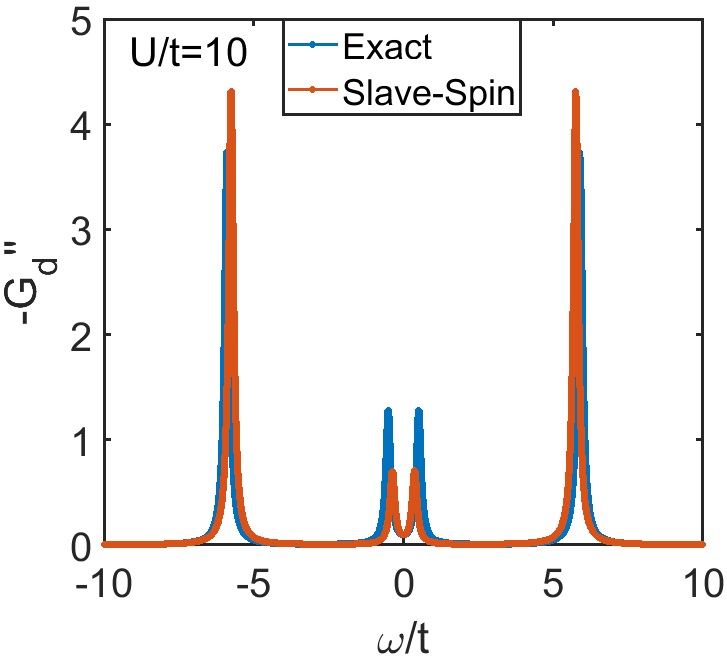}
\caption{\small A Comparison of the exact vs. slave-spin solutions of the single-orbital two-site problem for $U=10t$. The smaller hybridization gap and the larger Hubbard-gap is seen in the spectra. For $U\ll t$ and $U\gg t$ the two solutions agree much better than the intermediate regime shown here.}\label{fig:fig7}
\end{figure}

Fig.\,\pref{fig:fig7} shows a comparison of the exact solution to the one obtained from slave-spin method in the single-orbital two-site problem for $U=10t$. It can be shown that when $U/t\gg 1$ or $U/t\ll 1$ the two plots coincide.
\subsection*{2. Two-orbital problem}
In the two-orbital case, we can again diagonalize the Hamiltonian and the states are of the form $\ket{\alpha_1\alpha_2,\beta_1,\beta_2}$. Assuming $t_2=0$, the state $\ket{\beta_2}$ factors outs and we can drop it out. The remaining states depend if $\alpha_2=\ua,\da$ or $\alpha_2=0,2$. In the former case, again $\ket{\alpha_2}$ factors out (do not mix):
\be
\forall \alpha_2=\ua,\da, \qquad \ket{\alpha_1\alpha_2,\beta_1}=\ket{\alpha_2}\ket{\alpha_1,\beta_1}
\ee
{since as long as $\alpha_2$ is singly-occupied, the interaction is blind to the spin of $\alpha_2$},
and we get again the representation of Fig.\,\pref{fig:fig6} for each $\alpha_2=\ua,\da$. However, for $\alpha_2=0,2$ the states mix and we find a new set of atomic states shown in Fig.\,\pref{fig:fig8}. Each block has distinct $Q^z$ and $S^z$ quantum numbers. 
\begin{figure}[h!]
\includegraphics[width=1\linewidth]{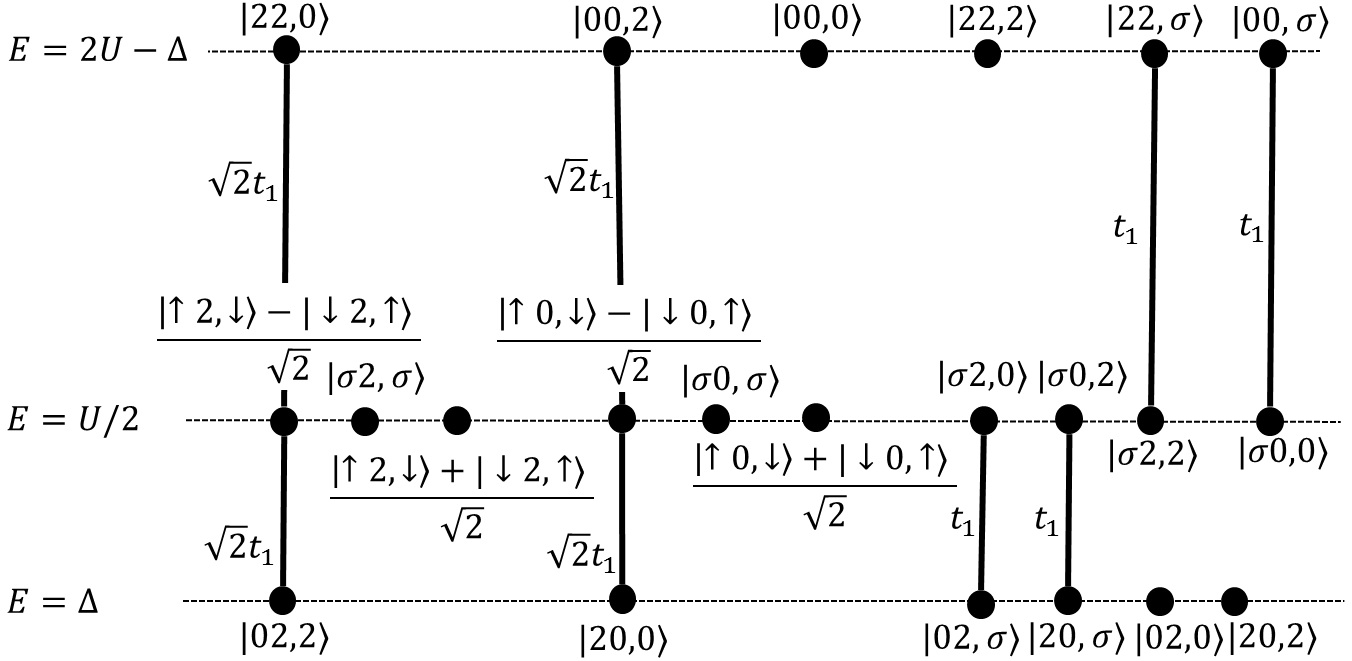}
\caption{\small A block-diagonal diagrammatic representation of the sectors of the Hamiltonian for two-orbital two-site problem, assuming $t_2=0$ and the second orbital, denoted by $\alpha_2$ in $\ket{\alpha_1\alpha_2,\beta_1}$ is restricted to empty or fully-occupied states $\alpha_2=0,2$ {due to the choice $t_2=0$. For each of the $\alpha_2=\ua,\da$ states, the Hamiltonian becomes a copy of single-orbital physics in Fig.\,8.} Again the black circles show the diagonal entries of the Hamiltonain matrix with the corresponding energy and the lines between the circles show the off-diagonal entries of the Hamiltonain. }\label{fig:fig8}
\end{figure}
The rightmost block of Fig.\,\pref{fig:fig6} has the lowest energy and is the ground state (degenerate due to $\alpha_2=\ua,\da$). The Green's function $G_{d1}(\tau)=\langle{-Td_1(\tau)d_1\dg}\rangle$ is, therefore, as calculated before. In order to calculate $G_{d2}(\tau)=\langle{-Td\dn_2(\tau)d\dg_2}\rangle$, we need to see which transitions are causes when $d_2\dg$ acts on the ground state. This is shown in Fig.\,\pref{fig:fig9} where the right side of Fig.\,\pref{fig:fig6} is shown in combination with the left side of Fig.\,\pref{fig:fig8}.
\begin{figure}[h!]
\includegraphics[width=\linewidth]{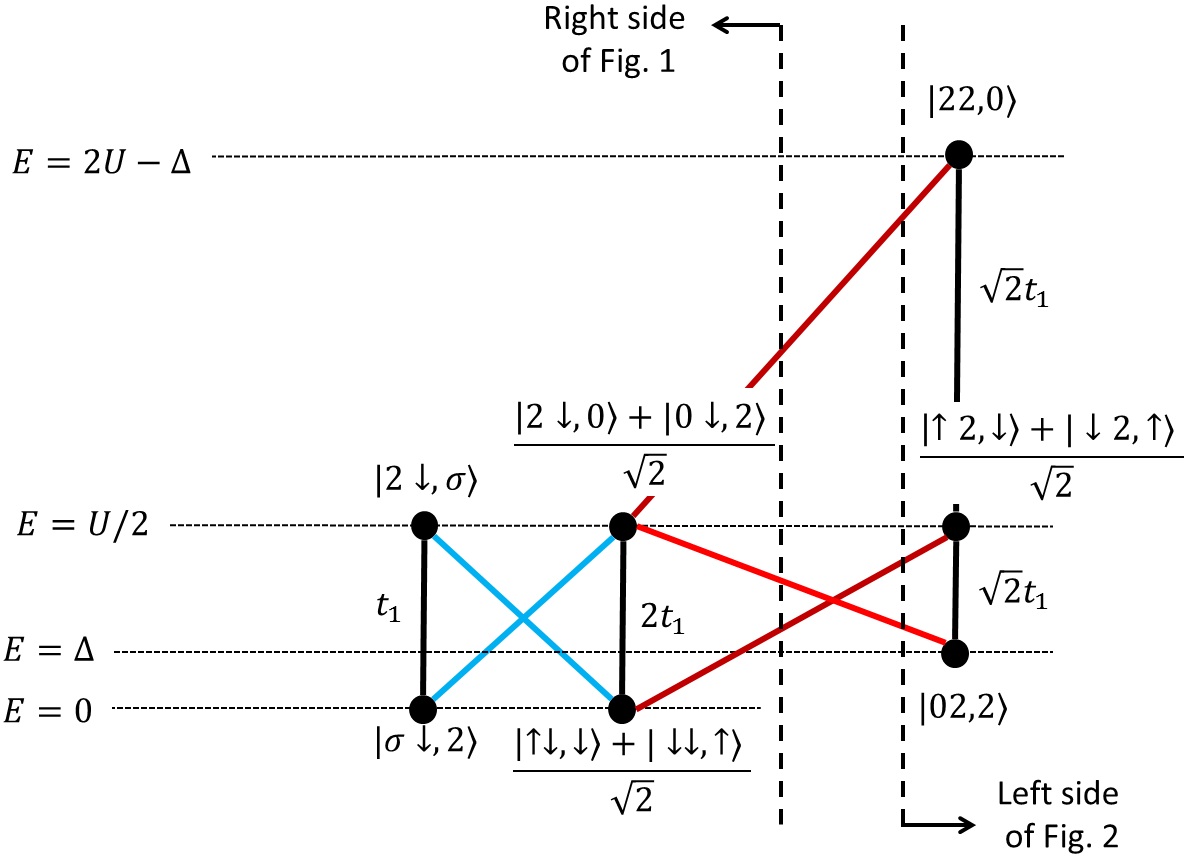}
\caption{\small A combination of Fig.\,8 and Fig.\,10 to highlight the ground state (middle) block for two-site problem. The creation operators $d_1\dg$ and $d_2\dg$ cause transitions marked in blue and red, respectively. See the main text.}\label{fig:fig9}
\end{figure}
Transitions caused by $d\dg_2$ to the states with a doublon at orbital 2 are shown in red color in this diagram. The transition caused by $d\dg_1$ follow the same blue arrows we had before, and therefore $G_1(\tau)$ is not modified. For $G_2(\tau)$ we have
\be
G_2(\tau>0)=-e^{-\beta\Omega}\braket{\psi_g\vert e^{\tau H}d\dn_{2\ua}e^{-\tau H}\bb 1 d\dg_{2\ua}\vert \psi_g}
\ee
The intermediate states appearing in  $\bb 1$ are indicated in the figure, have energies $E_n$ for $n=1,2,3$ and are of the form
\be
\ket{\psi_n}=\alpha_n\ket{02,2}+\beta_n\frac{\ket{\ua2,\da}+\ket{\da2,\ua}}{\sqrt 2}+\gamma_n\ket{22,0}\label{eqpsi_n}
\ee 
The parameters $\alpha_n,\beta_n,\gamma_n,E_n$ has to be determined by diagonalizing the corresponding $3\times 3$ matrix. As a result
\be
G_{2}({\rm z})=\frac{1}{2}\sum_n\Big[\frac{\abs{m_{ng}}^2}{{\rm z}-(E_n-E_g)}-({\rm z}\to-{\rm z})\Big],
\ee
Using Eqs.\,\pref{eqpsig} and \pref{eqpsi_n} we have
\be
m_{ng}=\braket{\psi_n\vert d\dg_{2\ua}\vert\psi_g}=\beta_n\cos\theta_g+\frac{\alpha_n+\gamma_n}{\sqrt 2}\sin\theta_g.
\ee
where $\theta_g$ determines the degree of the admixtures in the ground state.
\subsubsection*{Slave-spin}
The diagrammatic representation of the slave-spin Hamiltonian is shown in Fig.\,\pref{fig:fig3} with the states listed in Eq.\,\pref{eq25}. Most generally, at zero temperature the doublon part of $\Pi_2({\rm z})$ is given by
\ben
\Pi_2({\rm z})=\frac{\abs{m_{21,g}}^2}{{\rm z}-\Delta E_{21}}+\frac{\abs{m_{22,g}}^2}{{\rm z}-\Delta E_{22}}+\frac{\abs{m_{3,g}}^2}{{\rm z}-\Delta E_{3}}+({\rm z}\to-{\rm z})
\een
where $\Delta E_n\equiv E_n-E_g$. The lowest energy resonance is in the first term. Keeping only that term and writing both doublon and holon contribution we have
\be
\Pi({\rm z})=\abs{m_{+2,g}}^2\Big[\frac{1}{{\rm z}-\Delta E_{21}}-\frac{1}{{\rm z}+\Delta E_{21}}\Big]
\ee
\section*{Appendix B - Diagonalizing the Slave-spin Hamiltonian}
Using the notation
\be
\ket{\ua_1}=\ket{\Ua_{1\ua}\Da_{1\da}}, \quad \ket{2_1}=\ket{\Ua_{1\ua}\Ua_{1\da}}, \quad
\ket{0_1}=\ket{\Da_{1\ua}\Da_{1\da}},\nonumber
\ee
a simplified choice of basis for atomic states is given by
\bea
\ket{\psi_{11\pm}}&=&\frac{\ket{\ua_1}+\ket{\da_1}}{\sqrt{2}}\frac{\ket{\ua_2}\pm \ket{\da_2}}{\sqrt 2}\\
\ket{\psi_{13\pm}}&=&\frac{\ket{\ua_1}-\ket{\da_1}}{\sqrt 2}\frac{\ket{\ua_2}\mp \ket{\da_2}}{\sqrt 2}\\
\ket{\psi_{12\pm}}&=&\frac{\ket{2_1}\pm \ket{0_1}}{\sqrt 2}\frac{\ket{\ua_2}+\ket{\da_2}}{\sqrt 2}\\
\ket{\psi_{22\pm }}&=&\frac{\ket{\ua_1}+\ket{\da_1}}{\sqrt 2}\frac{\ket{2_2}\pm \ket{0_2}}{\sqrt 2}\\
\ket{\psi_{23\pm }}&=&\frac{\ket{2_1}\mp \ket{0_1}}{\sqrt 2}\frac{\ket{\ua_2}-\ket{\da_2}}{\sqrt 2}\\
\ket{\psi_{24\pm}}&=&\frac{\ket{\ua_1}-\ket{\da_1}}{\sqrt 2}\frac{\ket{2_2}\mp \ket{0_2}}{\sqrt 2}\\
\ket{\psi_{21\pm}}&=&\frac{\ket{2_1}\ket{0_2}\pm \ket{0_1}\ket{2_2}}{\sqrt 2}\\
\ket{\psi_{3\pm}}&=&\frac{\ket{2_1}\ket{2_2}\pm \ket{0_1}\ket{0_2}}{\sqrt 2}
\eea
and their energies are shown in the vertical axis in Fig.\,\pref{figdj1}. The transition caused by acting on these atomic states with $\tau^x_{1\sigma}$ and $\tau^x_{2\sigma}$ are indicated in blue/red, respectively with the indicated amplitudes. Since the kinetic Hamiltonian $a_r\tau^x_{r\sigma}$ has equal amplitudes for $\sigma=\ua,\da$, the dashed lines cancel each other and they drop out.
\begin{figure}[t!]
\includegraphics[width=\linewidth]{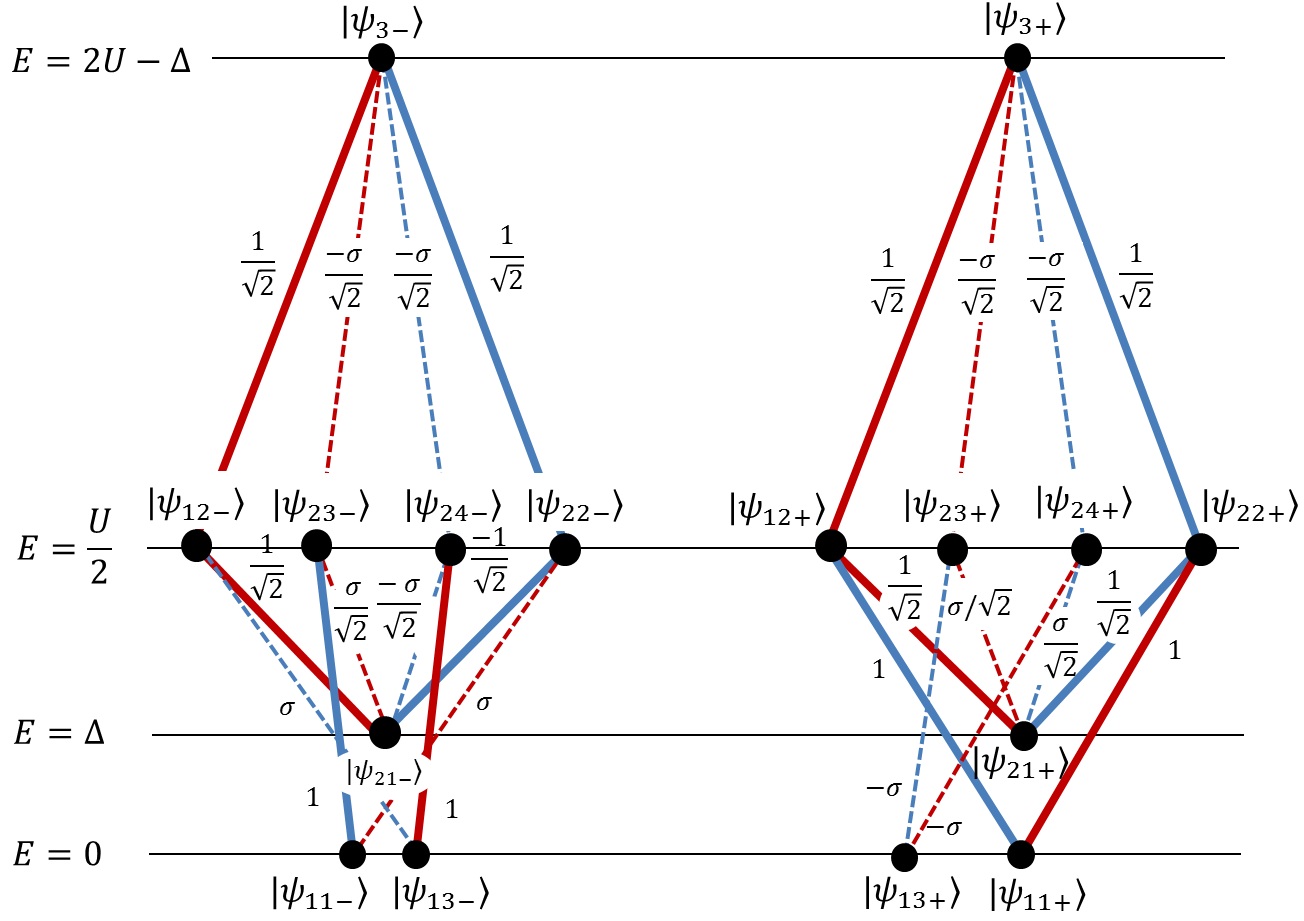}\caption{\small A representation of the atomic states of the salve-spin Hamiltonian. The blue/red lines are transitions caused by $\tau^x_{1\sigma}$ and $\tau^x_{2\sigma}$, respectively with amplitudes indicated. The dashed lines have an amplitude with a sign that depends on $\sigma$ and they all drop out in the paramagnetic phase of the Hamiltonian.}\label{figdj1}
\end{figure}

For the purpose of the paper note that when $t_2=0$ or $a_2=0$, all the red lines (as well as all the blue dashed lines) drop out and the Hamiltonian becomes a doubly degenerate (plus and minus sectors) version of Fig.\,\pref{fig:fig3}. Moreover, a chemical potential $\mu$ couples the two sector.

\section*{Appendix C - Spectral representation of $\bf G_d$}
Eq.\,\pref{eq6} in the real frequency reads
\be
G''_d(\omega)=-\int{\frac{dx}{\pi}}G''_f(\omega-x)\Pi''(x)[f(\omega-x)+n_B(-x)]\nonumber
\ee
where $G''(\omega)\equiv \im{G(\omega+i\eta}$.
This can be combined with Eq.\,\pref{eq8}, but to go to zero temperaure, we need to separate out the wavefunction normalization part. Assuming that the slave-spin Hamiltonian has a non-degenerate ground state, the resulting spectrum at zero temperature is
\bea
G''_d(\omega)&=&ZG''_f(\omega)\\
&&+\sum_{n\neq g}\Big[G''_f(\omega+\Delta E_n)\theta(-\omega> \Delta E_n)\abs{\braket{n\vert z_\alpha\vert g}}^2\nonumber\\
&&\hspace{0.8cm}+G''_f(\omega-\Delta E_n)\theta(\omega>\Delta E_n)\abs{\braket{g\vert z_\alpha\vert n}}^2\Big].\nonumber
\eea
Here, $\Delta E_n=E_n-E_g$ and the ground state is treated separately and excluded from the summation.
%\phantomsection	
%\addcontentsline{toc}{chapter}{Bibliography}
\bibliography{slave} %bibliography file with bibtex
%-------------------------------------------------------------------
\end{document}